\documentstyle[twocolumn,prl,aps]{revtex}
\begin{document}

\title{
Pinning of quantized vortices in helium drops  \\
by dopant atoms and molecules
}

\author{Franco Dalfovo}

\address{
Dipartimento di Matematica e Fisica, Universit\`a Cattolica, Via 
Trieste 17, I-25121 Brescia, Italy
\\
and 
Istituto Nazionale per la Fisica della Materia, Unit\`a di Trento,
I-25121 Povo, Italy}

\author{Ricardo Mayol, Mart\'{\i} Pi, and Manuel Barranco}

\address{
Departament ECM,  Facultat de F\'{\i}sica,
Universitat de Barcelona.
E-08028 Barcelona, Spain}

\date{22 December 1999}

\maketitle

\begin{abstract}

Using a density functional method, we investigate the properties of 
liquid $^4$He droplets doped with atoms (Ne and Xe) and molecules 
(SF$_6$ and HCN). We consider the case of droplets having a quantized 
vortex pinned to the dopant. A liquid drop formula is proposed that 
accurately describes the total energy of the complex and allows one 
to extrapolate the density functional results to large $N$. For a given 
impurity, we find that the formation of a dopant+vortex+$^4$He$_N$ 
complex is energetically favored below a critical size $N_{\rm cr}$. 
Our result support the possibility to observe quantized vortices in 
helium droplets by means of spectroscopic techniques. 
\end{abstract}


Since the first observation of the $\nu_3$ vibrational band of SF$_6$ 
dissolved in $^4$He droplets \cite{Goy92}, the infrared  
spectroscopy of molecules inside or attached to helium
has attracted a wide interest (see, for instance, 
Refs.~\cite{Har95,Wha98,Toe98} and references therein). A major 
motivation for these efforts is that cold helium droplets offer the 
possibility of resolving rotational spectra of rather complex molecules 
and may constitute `the ultimate  spectroscopic matrix' \cite{Leh98} 
to create and study novel species. This unique feature of helium
droplets originates from their quantum nature: not only they are 
fluid at zero temperature, due to the large zero point motion, but 
also exhibit a crucial superfluid behavior.
The superfluid character of $^4$He droplets is interesting also from a
fundamental viewpoint. In fact, the observation of superfluid effects in 
finite-sized quantum systems has to do with important concepts, like 
order parameter, Bose-Einstein condensation, and phase coherence, which
were originally introduced for uniform systems and which are now widely
used in different contexts.

In the case of liquid helium, Grebenev {\it et al.} \cite{Gre98} 
recently showed that only a rather small amount of $^4$He atoms is
needed to develop a superfluid droplet, also confirming theoretical 
predictions \cite{Sin89}. In that experiment, the evidence for 
superfluidity is the appearence of a sharp rotational spectrum of an 
OCS molecule in $^3$He-$^4$He mixed drops, when the number of $^4$He 
atoms surrounding the dopant \cite{Pi99} is larger than about $60$.   
In the same spirit, experiments have been made to observe critical
velocities \cite{Har96} (i.e., the occurrence of a Landau criterion for 
superfluidity), and a reduction of the moment of inertia (see
\cite{Cal99} and references therein). In contradistinction,  
detecting quantized vortices in droplets still remains an
open question. It is worth stressing that all these investigations
have many analogies with the current activity on Bose-Einstein 
condensation in trapped gases, where new results are now available
about critical velocities \cite{Ram99}, moments of inertia
\cite{Str96,Mar99} and vortices \cite{Mat99}. 

In this work we address the problem of quantized vortices. One
first observes that a vortex line in a pure droplet is expected 
to be difficult to produce and stabilize,  since it implies a 
significant increase of energy compared to a vortex-free droplet.
In order to circumvent this limitation, we explore the possibility 
of pinning the vortex line to a dopant atom or molecule. If the 
dopant is deeply bound inside the droplet, it might stabilize the 
vortex for a time long enough to permit its observation.
A second advantage is that the dopant could make the detection 
feasible {\it via} spectroscopic techniques.  

Our purpose is to determine the energy and density profile of a
impurity+vortex+$^4$He$_N$ complex,  for droplets up to $N$= 1000, 
using a finite-range  density functional.
We then subtract to its energy  that of the same  droplet without
vortex and/or impurity and  show that the difference fits very
well to a liquid drop formula, which allows one to safely 
extrapolate to larger droplets. The density functional method 
consists in minimizing the  total energy of the system at zero 
temperature written as a functional of the He density. We use the
Orsay-Paris functional \cite{Dup90}, which is based on an effective 
non-local interaction with a few parameters fixed to reproduce known 
properties of bulk liquid He. This functional has been shown
to accurately reproduce the static properties of pure and 
doped He clusters~\cite{Dal94} and has also been used to describe a
quantized vortex line in bulk liquid helium\cite{Dal92}. In the
latter case, the vortex is included with the Feynman-Onsager (OF)
{\it ansatz} for the velocity field. This implies  a singular vorticity 
and hence the vanishing of the density on the vortex axis. At $T$= 0 
this approximation is a reasonable starting point, since
it enormously reduces the computational cost. Recent calculations 
\cite{Sad97} have shown that the density profile and energy of the 
vortex line given by the OF approximation are reasonably close to 
the ones obtained by assuming non-singular vorticity. Finally, 
the actual temperature of the droplets, below $0.4$ K \cite{Har95}, is 
low enough for neglecting thermal contributions to the fluid motion. 
Thus, corrections beyond OF and at $T\neq 0$ are expected not to 
change the main results of the present work.

The minimization of the energy is performed in axial symmetry 
by mapping the density on a grid of points, putting the vortex line
along the $z$-axis and the dopant in the center, at ${\bf r}=0$. The 
numerical code used to calculate the density profile and energy is the 
same used in  \cite{Cal99}.  The potentials for rare gas impurities have
been taken from \cite{Tan86}, that of the spherically averaged SF$_6$ 
from \cite{Pac84}, and that of HCN from \cite{Atk96}.

We first consider pure droplets with and without vortex. In
Fig.~\ref{fig:profiles} we show density profiles, at $z=0$,
obtained for different $N$. For large droplets the shape approaches 
that of a rectilinear vortex in the uniform liquid \cite{Dal92};
the core radius is of the order of $1 \div 2$ \AA \  and the density
oscillates as a consequence of the He-He interaction. In 
Fig.~\ref{fig:deltaev} we plot the energy associated to the
vortex flow, defined as
\begin{equation}
\Delta E_{\rm V} (N) \equiv E_{\rm V}(N) -E(N) \; ,
\label{eq:deltaev}
\end{equation}
where $E_{\rm V}$ and $E$ are the energies of droplets with and 
without vortex, respectively. The solid line represents the results
obtained with a liquid drop formula
of the kind:
\begin{equation}
\Delta E_{\rm V} (N) = \alpha N^{1/3} + \beta N^{1/3} \log N + 
\gamma N^{-1/3}  \;  ,
\label{eq:ld-deltaev}
\end{equation}
with the parameters $\alpha$= 2.868 K, $\beta$= 1.445 K, and
$\gamma$= 0.313 K extracted from a fit to the density functional 
calculations. This formula works well. The reason can be easily 
understood
by means of a hollow-core model for the vortex, having core radius 
$a$, in a droplet of radius $R$ and constant density $\rho_0$. By 
integrating the kinetic energy of the vortex flow in the limit 
$R\gg a$, one gets 
\begin{equation}
E_{\rm kin}  = \frac{2\pi\hbar^2\rho_0}{m_4}
\left[R \log\left(\frac{2 R}{a}\right) - R +
\frac{a^2}{4R}\right] \; .
\label{eq:hollow}
\end{equation}
Writing $R$= $r_0 N^{1/3}$ one recovers the $N$-dependence as in 
(\ref{eq:ld-deltaev}).

The next step is the inclusion of a dopant atom or molecule. As an example,
in Fig.~\ref{fig:hcn} we show the He density distribution for a drop of 
$N = 500$ with HCN hosted in the vortex core. Both the  axis of the linear 
molecule and that of the vortex are taken along $z$. The density is
very inhomogeneous near the dopant, due to the complexity of the 
HCN-He interaction. The energetics of the system can be conveniently
analysed by introducing the following energies:  
\begin{eqnarray}
\Delta E_{\rm V}^{\rm X} (N) 
& \equiv & E_{\rm X+V} (N) - E_{\rm X} (N) \; ,
\label{eq:solvation1} \\
S_{\rm X} (N) & \equiv & E_{\rm X} (N) -E (N)  \; ,
\label{eq:solvation2} \\
S_{\rm X+V}(N) & \equiv &E_{\rm X+V} (N) -E (N) \; ,
\label{eq:solvation3}
\end{eqnarray}
where the subscripts $X$ and $V$ refer to drops doped with impurity
$X$ and/or vortex line. 

The energy $\Delta E_{\rm V}^{\rm X}$ is the one associated with the 
vortex flow in the doped cluster. In Fig.~\ref{fig:deltaev} it is 
compared with the vortex energy in pure droplets, $\Delta E_{\rm V}$.
The difference 
\begin{equation}
\delta_{\rm X} (N) = 
\Delta E_{\rm V}^{\rm X} - \Delta E_{\rm V}  < 0
\label{eq:delta}
\end{equation}
is almost independent of $N$, apart from the smallest droplets. The
reason is that this difference has to do with the `geometrical extension' 
of the dopant, i.e., the `hole' made by the dopant in the vortex flow, 
as well as with the distortion of the density near the dopant caused
by the pinning of the vortex core. Both effects are localized near
the dopant and, thus, they are expected to give a shift in energy 
which becomes $N$-independent for large droplets.

The quantity $S_{\rm X} (N)$ in Eq.~(\ref{eq:solvation2}) is the 
solvation energy of the dopant in a vortex-free droplet. 
The results obtained for Ne, Xe, HCN and SF$_6$ are 
shown in Fig.~\ref{fig:sx}. As already discussed in \cite{Dal94,Gat97},
the solvation energy becomes almost $N$-independent for $N$ larger
than a few hundreds. The value at $N=1000$ can be safely taken 
to represent the solvation energy in the bulk, $S_{\rm X} (\infty)
\simeq S_{\rm X} (1000)$.  For our analysis, we have chosen 
impurities having binding energies on a wide range. 

The key quantity in the present study is the solvation energy
of the dopant+vortex complex  given by $S_{\rm X+V} (N)$ in 
Eq.~(\ref{eq:solvation3}). The results are shown in Fig.~\ref{fig:sxv}.
 From the definitions (\ref{eq:solvation1})-(\ref{eq:delta}) one can
also write 
\begin{eqnarray}
S_{\rm X+V}(N) & = & E_{\rm X} (N) + 
\Delta E_{\rm V}^{\rm X} (N) -E (N) 
\nonumber \\
& = & S_{\rm X} (N) + \Delta E_{\rm V} (N) + \delta_{\rm X} (N) \; . 
\label{eq:sum}
\end{eqnarray}
In Fig.~\ref{fig:sxv} we compare $S_{\rm X+V}$ with the sum 
$ S_{\rm X}+ \Delta E_{\rm V}$; the difference is $\delta_{\rm X}$.
The simple picture which emerges from this analysis is that the 
solvation energy of the dopant+vortex complex is just the sum of 
the solvation energy of the dopant with no  vortex and the extra 
energy of a vortex in a pure droplet, apart from a small shift 
which depends on the dopant. Deviations from this rule are 
significant only for small droplets, having radius
of the order of the size of the dopant. Our numerical results 
provide a quantitative basis for this picture and yield typical
estimates of $\delta_{\rm X}$. It is worth noticing that, by 
rearranging the terms in (\ref{eq:delta}), this quantity can 
be written as the difference between the solvation energies of the 
dopant in a droplet with an without vortex, $\delta_{\rm X} =
[(E_{\rm X+V}-E_{\rm V}) - (E_{\rm X} -E)]$, and can hence be  
interpreted as the binding energy of the dopant to the 
vortex \cite{Sadd99}.  

Since the solvation energy $S_{\rm X}$ is negative and almost 
constant for $N > 300$ while the vortex energy $\Delta E_{\rm V}$
always increases, the dopant+vortex complex has a solvation energy
which changes sign at some $N_{\rm cr}$. This means that for 
$N<N_{\rm cr}$, the dopant+vortex complex is energetically favored.  
In the case of Ne, as one can see in
Fig.~\ref{fig:sxv},  $N_{\rm cr} \sim 380$. This
number is rather small as compared to the typical droplet size
in current experiments, and is a consequence of the weak binding
of Ne. Dopants with stronger binding have larger $N_{\rm cr}$.
An estimate for HCN, Xe, and SF$_6$ can be easily obtained by
means of the liquid drop formula. 
One has to insert expression (\ref{eq:ld-deltaev})
in (\ref{eq:sum}) and use the large-$N$ values of $S_{\rm X}$ and
$\delta_{\rm X}$. The $S_{\rm X}$ values turn out to be
 $-310$ K, $-320$ K and $-622$ K,  and the $\delta_{\rm X}$ 
values are  5.0 K, 4.4 K, and 7.7 K,
for Xe, HCN, and SF$_6$, respectively. These numbers yield
$N_{\rm cr} \simeq 7600$ for Xe, $\simeq 8100$ for HCN, and $\simeq
40000$ for SF$_6$. 

In conclusion, the analysis of the energetics of doped helium 
droplets has allowed us to disclose a possible mechanism to create 
and stabilize vortex lines. A dopant+vortex complex could be formed 
by picking up the impurity, assuming that the collision imparts
sufficient angular momentum. The vortex line is expected to appear 
attached to the dopant, since the binding energy $\delta_{\rm X}$ is 
negative.  The formation of the complex is energetically favored 
below a critical $N$ which is well within
the range of droplet sizes met in current experiments  
if the dopant has a large solvation energy. A metastable state could 
also exists for $N > N_{\rm cr}$, but  estimating  its lifetime is 
a more demanding calculation. One should also explore the energy 
barrier associated with other possible decay processes. Further work 
is planned in this direction. 

We thank Kevin Lehmann for useful discussions. F.D. would like to thank
the Dipartimento di Fisica, of the Universit\`a di Trento, where
part of this work has been done. This work has been performed under
grants No. PB98-1247 from DGESIC, Spain, and No. 1998SGR-00011
from Generalitat of Catalunya.

\begin{figure}
\caption{ He density profiles (solid lines) 
in the $z=0$ plane of drops with $N$= 40, 100, 200, 300, 
400, 500, and 1000 having a vortex line along the $z$-axis. 
The density profiles of vortex-free droplets are also shown 
(dashed lines). 
}
\label{fig:profiles}
\end{figure}
\begin{figure}
\caption{ The vortex energy $\Delta E_{\rm V} (N)$ (dots) in 
pure $^4$He$_N$ drops. The line is a fit obtained using 
Eq.~(\protect\ref{eq:ld-deltaev}). Open symbols are the vortex 
energies $\Delta E_{\rm V}^{\rm X} (N)$ in doped droplets. 
}
\label{fig:deltaev}
\end{figure}
\begin{figure}
\caption{ Density distribution in the $xz$-plane for a 
HCN-He$_{500}$ droplet hosting a vortex along the z-axis.
Lengths are in {\rm \AA}.
Contour lines are drawn bewteen $10$ equally spaced intervals
of density,  where white is for density less than $5\times 
10^{-3}$~\AA$^{-3}$ and black for density higher than 
$5\times 10^{-2}$~\AA$^{-3}$. 
}
\label{fig:hcn}
\end{figure}
\begin{figure}
\caption{Solvation energy $S_{\rm X} (N)$ 
for $X$= Ne, Xe, SF$_6$, and HCN. The lines have been 
drawn to guide the eye. 
}
\label{fig:sx}
\end{figure}
\begin{figure}
\caption{Excess energies $S_{\rm X+V}(N)$ (dots) 
for Ne, Xe, SF$_6$, and HCN. The triangles represent 
$S_{\rm X} (N)+\Delta E_{\rm V} (N)$. Lines have been drawn 
to guide the eye. 
}
\label{fig:sxv}
\end{figure}

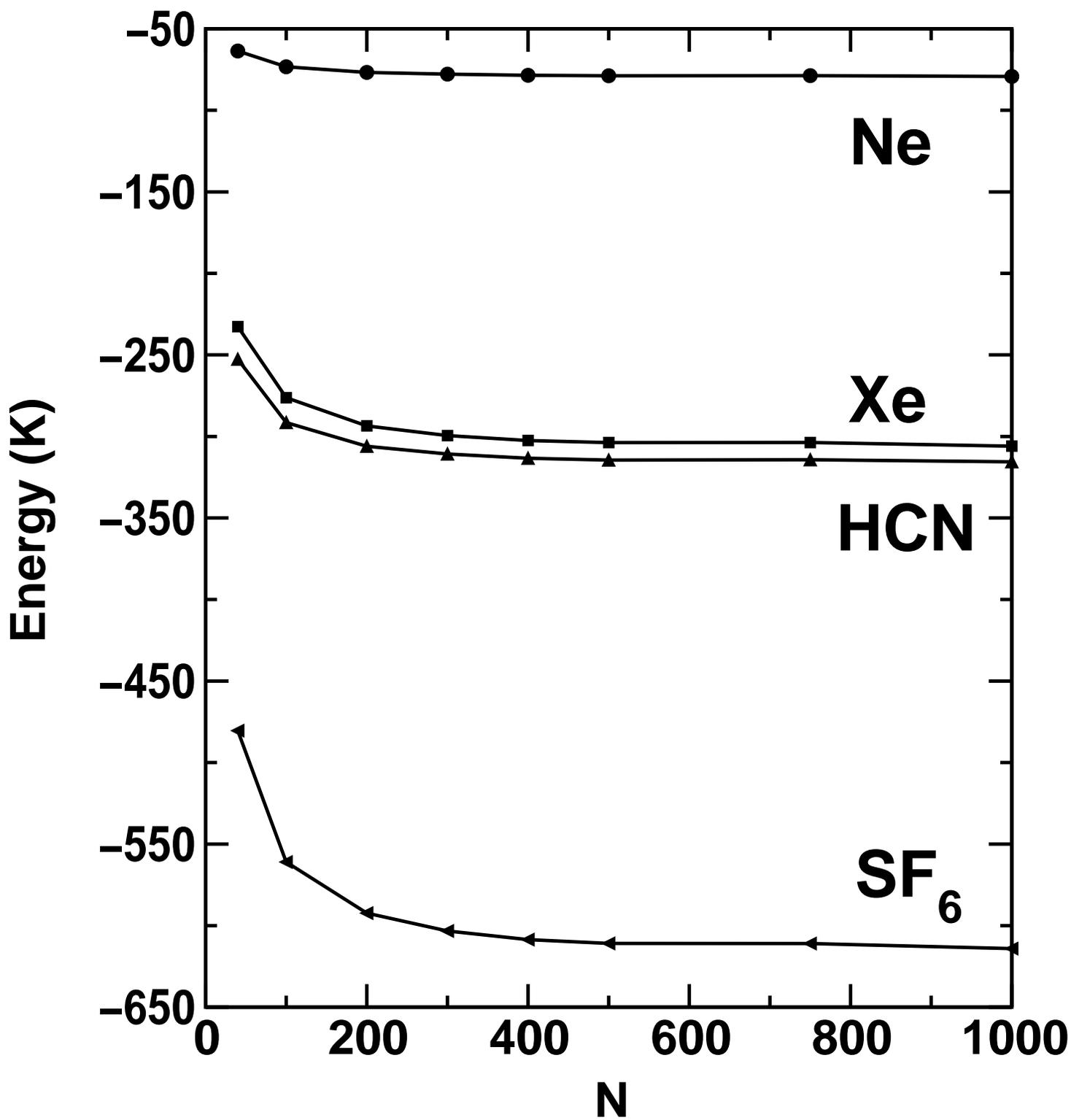
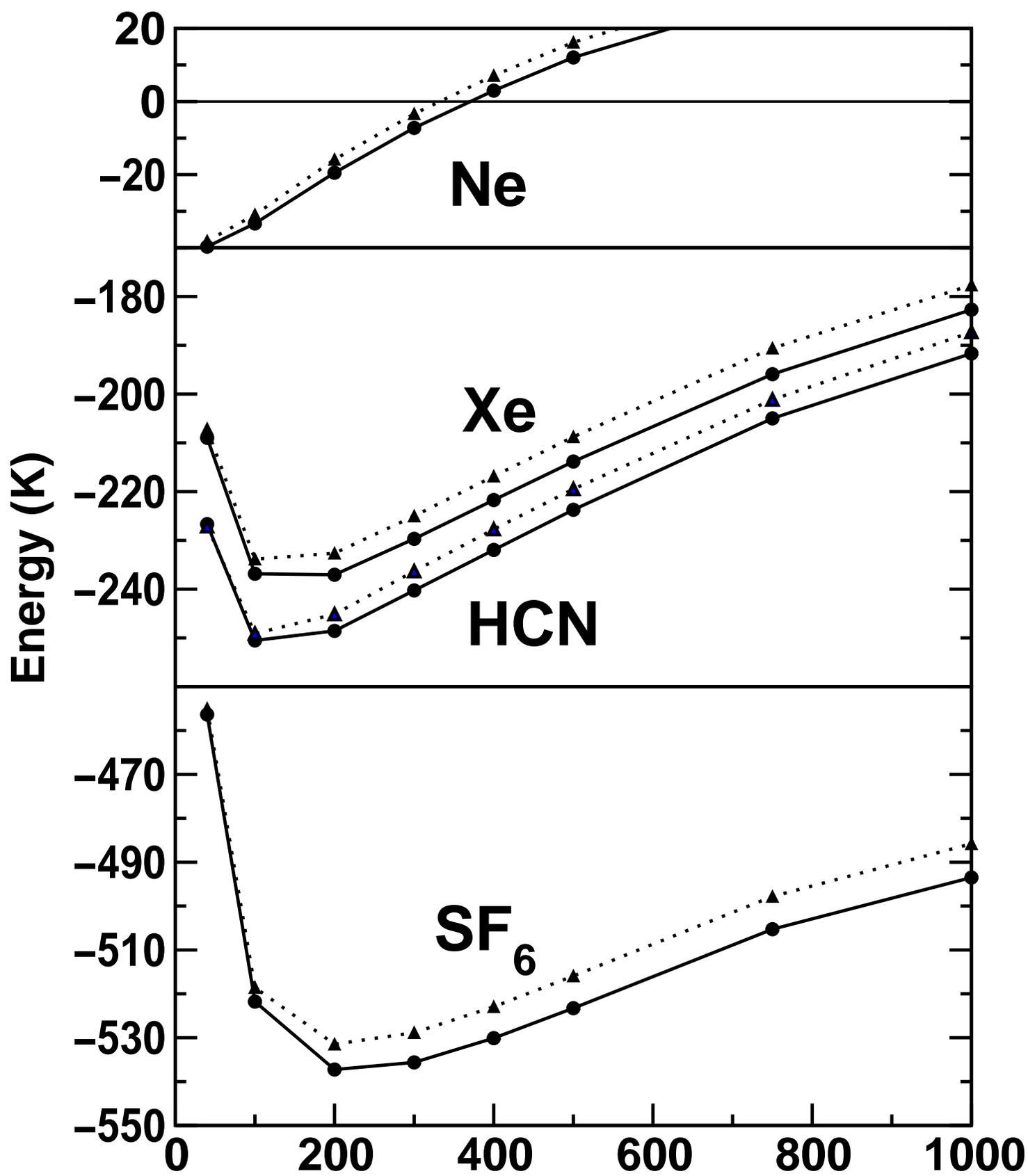
\end{document}